\documentclass[10pt,%showpacs,nofootinbib,
         amsmath,
         amssymb,
         aps,
         prl, 
         longbibliography,
         floatfix,
         twocolumn
]{revtex4-2}
\usepackage[english]{babel}
\usepackage[utf8]{inputenc} 
\usepackage{dsfont}
\usepackage{SIunits}
\usepackage[caption=false,label font=bf,labelformat=simple]{subfig}
\usepackage{graphicx}     % Allows to insert images from outside
\usepackage{color}
\usepackage[normalem]{ulem}
\usepackage{soul}
\usepackage{cancel}
\usepackage{hyperref}% add hypertext capabilities

\begin{document}

\newcommand{\red}[1]{\textcolor{red}{#1} }
\newcommand{\blue}[1]{\textcolor{blue}{#1} }
\newcommand{\delb}[1]{\textcolor{blue}{\sout{#1}}}
\newcommand{\del}[1]{\textcolor{red}{\sout{#1}}}
\newcommand{\bfs}[1]{\boldsymbol{#1}}

\newcommand{\keldysh}[1]{\left\langle \left\{T_K #1  \right\}\right\rangle}
\newcommand{\sq}[1]{\left[#1\right]}
\newcommand{\crl}[1]{\left\{#1\right\}}
\newcommand{\rnd}[1]{\left(#1\right)}
\newcommand{\ang}[1]{\left<#1\right>}
\newcommand{\charge}{q_{\alpha}}
\newcommand{\scale}{\delta_{{\alpha}}}
\newcommand{\ampl}{\Gamma_{\alpha}}
\newcommand{\chargesing}{q_{\boldsymbol{g}}}
\newcommand{\scalesing}{\delta}
\newcommand{\amplsing}{\Gamma_{\boldsymbol{g}}}

\definecolor{Jerome}{rgb}{0.8, 0.0, 0.3}
\newcommand {\jr} {\textcolor{Jerome}}
\definecolor{Flavio}{rgb}{0.1, 0.9, 0.1}
\newcommand {\fr} {\textcolor{Flavio}}
\definecolor{Thibaut}{rgb}{0.2, 0.0, 0.8}
\newcommand {\tj} {\textcolor{Thibaut}}
\newcommand{\deltj}[1]{\textcolor{Thibaut}{\sout{#1}}}

\title{Photo-assisted shot noise probes multiple charge carriers in quantum Hall edges}

\author{K. Iyer}
\email{kishore.iyer@etu.univ-amu.fr}
\author{F. Ronetti}
\author{B. Gr\'emaud}
\author{T. Martin}
\author{T. Jonckheere}
\author{J. Rech}
\affiliation{Aix Marseille Univ, Université de Toulon, CNRS, CPT, Marseille, France}%

\date{\today}

\begin{abstract}
Fractional charges in the fractional quantum Hall effect were first observed via DC shot noise measurements of anyons tunneling at a quantum point contact (QPC). However, in scenarios with simultaneous tunneling of different types of charges at the QPC, the connection between DC shot noise and tunneling charge is less transparent. Photo-assisted shot noise (PASN), induced by periodic AC voltage, offers a promising alternative. Here, we investigate PASN in the hierarchical states of the fractional quantum Hall effect, where different types of charges are expected to tunnel concurrently at QPCs. In the particular case of the fractional quantum Hall state $\nu = 2/3$, our analysis demonstrates that PASN can be employed as a robust tool to detect different tunneling charges, even when the tunneling amplitude of one type is significantly smaller compared to the other. We show that the features predicted by our calculations are still visible for typical values of temperature and frequency achieved in state-of-the-art experiments. Our general formalism can be used to compute PASN for general Abelian quantum Hall systems with multiple edge modes and charge types.

\end{abstract}

\maketitle

\emph{Introduction.--} Fractional quantum Hall effect (FQHE) \cite{laughlin83,tsuiNobel} is predicted to host anyons, particles intermediate between fermions and bosons \cite{Leinaas1977,wilczek82}. These particles are characterized by striking properties such as fractional charge and fractional exchange phase. The past decades have witnessed intense and fruitful theoretical and experimental efforts in understanding these properties \cite{chamonFabryPerot,safi01,martin_lesHouches,rosenow16,bartolomei20,  nakamura20,Feldman_2021,interferometry_anyon,ruelle23,glidic23b,lee23braid,nakamura23,iyer24,thamm24, samuelson2024anyon,werkmeister2024anyon}.

The fractional charge of anyons has been measured three decades ago via the Fano factor, i.e. the ratio of DC shot noise and tunneling current \cite{kane94}, in a quantum point contact (QPC) geometry~\cite{saminadayar97,dePicciotto97, reznikov}. Since then, it has been confirmed through a variety of independent methods, such as finite-frequency noise \cite{safi07, Carrega2012, safi16, Ferraro2012, bisognin19} and photoassisted shot noise \cite{fqhepasn1, qhepasn1,qhepasn2,qhepasn3, safi22, kapfer19}. However, the most compelling agreement between experiments and theory based on chiral Luttinger liquids \cite{wenCLLPRB,wenCLLPRL}  is restricted to Laughlin fractions.

The correspondence between theory and experiments becomes more complex for states outside the Laughlin sequence \cite{halperin83,jain89,wen95review}. These states are characterized by richer edge structures and multiple types of anyons with distinct topological properties, including fractional charges, that could all tunnel simultaneously at a QPC \cite{shtanko14}. For instance, DC shot noise experiments have witnessed a temperature and QPC transmission dependence of the Fano factor \cite{ferraro_charge_2010,bid09charge,ferraro_relevance_2008,chung_scattering_2003,transmission1,biswas22,schiller24,Glidic2023,kane04,iyer23Andreev}. This inspired previous works to investigate multiple charges tunneling in non-Laughlin states focusing on other quantities, specifically the finite-frequency noise in the low-temperature regime~\cite{carrega12,Ferraro2014}. Since experimental evidence for these different charges is inconclusive~\cite{biswas22}, it is desirable to identify complementary probes that can robustly detect different charges tunneling simultaneously at the QPC.

\iffalse

~\cite{ferraro_charge_2010,bid09charge,ferraro_relevance_2008,chung_scattering_2003,Ferraro_2012,transmission1,transmission2}. Non-universal factors, such as the scaling dimension and tunneling amplitudes of quasiparticles, reduce the effectiveness of the Fano factor with respect to the case of states in the Laughlin sequence~\cite{biswas22,schiller24,Glidic2023,kane04,iyer23Andreev}. This emphasizes the need for experimental quantities that can clearly identify topologically protected fractional charges at the QPC, independent of non-universal details.
\fi

A groundbreaking experiment \cite{kapfer19} showed that superimposing an AC voltage at frequency $\Omega$ and a DC voltage $V_1$, the Josephson frequency of anyons, $\Omega_J = e^* V_1/\hbar$, can be extracted from photo-assisted shot noise (PASN) measurements. PASN, used in normal diffusive, ballistic systems~\cite{normal1,normal2,normal3} and superconductors~\cite{super1,super2}, measures current-current correlations in a quantum point contact (QPC) biased by both DC and AC voltages~\cite{qhepasn1,qhepasn2,qhepasn3}. In Ref. \cite{kapfer19}, the anyon Josephson frequency was observed by varying $V_1$, leading to resonances when
$\Omega_J=e^* V_{1}/\hbar$ equals an integer multiple of the driving frequency $\Omega$. The value of the fractional charge $e^*$ can then be extracted from $\Omega_J$. These discrete resonances render PASN a promising method for detecting fractional charges with potential interesting applications when multiple charges are involved.

In this Letter, we study PASN for a general Abelian, non-Laughlin state of the FQHE, where multiple edge modes are present, and different anyons with different charges can tunnel simultaneously at the QPC. For the sake of simplicity and clarity, we detail the case of $\nu=2/3$ in the main text, while the general formalism is exposed in the End Matter. Using chiral Luttinger liquid description for $\nu=2/3$, the edge is described by one charged boson mode and a counterpropagating neutral boson mode \cite{KFP}. At the QPC, three types of anyons can tunnel: two with charge $e/3$ and one with charge $2e/3$ \cite{shtanko14}. Following the approach of Refs. \cite{fqhepasn1, kapfer19}, we analyze PASN as a function of the DC voltage $V_1$. We find PASN minima at the Josephson frequencies $2e V_1/\hbar$ and $e V_1/\hbar$, reflecting the two different charges tunneling in the system. The resonances associated with the Josephson frequency of different anyons in PASN remain distinct and visible even at finite temperatures and for renormalized scaling dimensions $\delta$ (up to $\delta \sim 1$), being therefore independent of non-universal factors. Our method can hence be reliably employed to unambiguously detect the presence of different charges in the system and extract their values. 
%{need to say something about DC Fano factor ?}
%In the End Matter, we present a general theory for all Abelian FQHE edge states under combined AC and DC voltage drives and the general expression for the PASN, thus opening the way to generalization to other fractional states.

\begin{figure}[t]
    \centering
    \includegraphics[width=0.95\linewidth]{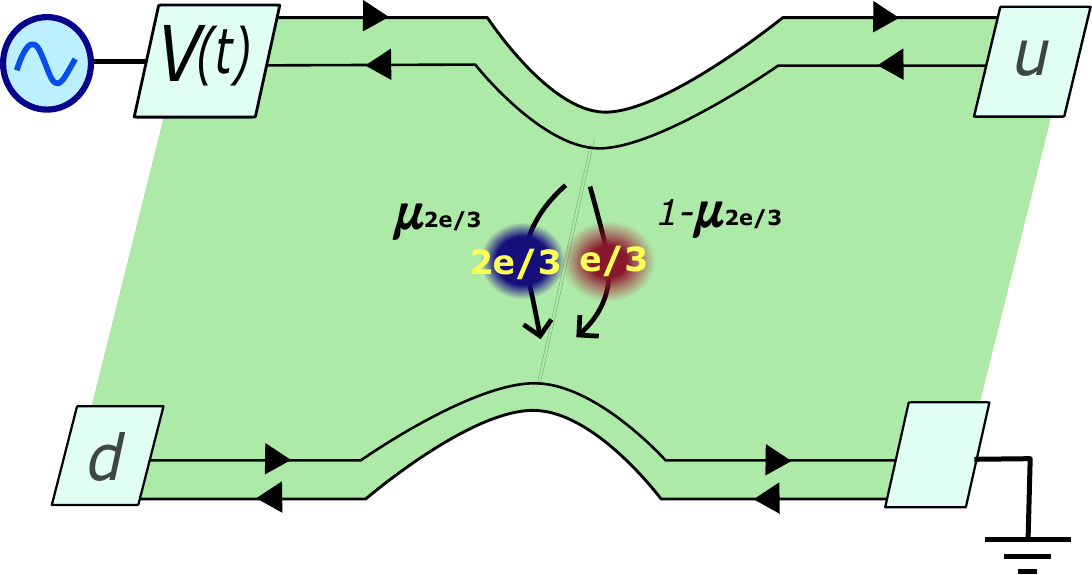}\caption{Schematic of the proposed setup. A quantum Hall system, endowed with a QPC in the anyon tunneling regime, is biased with a time-dependent voltage $V(t) = V_1 + V_2\sin(\Omega t).$ Different types of anyons can tunnel simultaneously across the QPC.}   
    \label{fig:Setup}
\end{figure}

\emph{Theoretical model.--}
Our setup is presented in Fig.~\ref{fig:Setup}. Each of the two FQHE edges at $\nu = 2/3$ is described by the Hamiltonian $H_0 = \sum_{j = 1}^{2} \frac{v_j}{4\pi} (\partial_x\phi_j)^2$\footnote{We set $\hbar = 1$ in the rest of the paper.}, where $\phi_1$ denotes the bosonic charge mode, $\phi_2$ the bosonic neutral mode, $v_{j} > 0$ denote the propagation velocity of $j-$th bosonic mode~\cite{Kane1994}. The bosonic fields satisfy commutation relations $   \left[\phi_j(x), \phi_k(y) \right] = i\pi\chi_j\delta_{jk}\text{sign}(x-y)
$, where $\chi_{1/2} = \pm 1$ is the chirality of the charge/neutral mode. The edges host anyons of the form $   \psi_{\alpha} = e^{i {\boldsymbol{g}_\alpha}.\boldsymbol{\phi}}$ where the vectors $\boldsymbol{g}_\alpha$, $\boldsymbol{\phi}$ with two components are denoted in bold. The corresponding parameters for each type of anyon are shown in Table~\ref{table:two-thirds_parameters}. We notice that they have different charges $q_{1} = 2e/3$, $q_{2} = q_{3} = e/3$, and the same scaling dimensions $\delta_{1} = \delta_{2} = \delta_{3} \equiv \scalesing = 2/3$. 

The two opposite edges of the FQHE bar are coupled by a point-like QPC, which is assumed to be tuned to the weak-backscattering regime. Since all scaling dimensions are identical, each type of anyon simultaneously tunnels across the QPC. The corresponding tunneling Hamiltonian is $H_T(t) = \sum_{\alpha }  \ampl e^{-i\omega_{\alpha}(t)} \psi^{\dagger d}_{\alpha}(0,t)\psi^{u}_{\alpha} (0,t)+ H.c.$ where $\ampl$ is the tunneling amplitude of type$-\alpha$ anyon, $e^{-i\omega_{\alpha}(t)}$ is the phase gathered by a type$-\alpha$ anyon as it tunnels across the QPC, and the superscripts $u/d$ denote operators on the upper/lower edge. 

The upper edge is driven by a time-dependent, periodic voltage  $V(t) = V_1 + V_2 \sin\left(\Omega t\right) $. A sinusoidal voltage is chosen for its simplicity and its relevance to recent experiments \cite{kapfer19}. In the presence of the driving voltage, the anyons acquire a phase $\exp\left[-i\omega_{\alpha}(t)\right] = \exp\left[-i\charge\int_{-\infty}^{t}dt'~V(t')\right]$~\cite{rech17, Ronetti24}. The latter can be conveniently recast as a discrete sum
\begin{equation}
    \exp\left[-i\omega_{\alpha}(t)\right] = \sum_{l \in \mathbb{Z}} J_{l}\!\rnd{\frac{\charge V_2}{\Omega}} e^{-il \Omega t}e^{-i q_{\alpha}V_1 t},
    \label{eq:floquet}
\end{equation}
where $J_{l}$ are the Bessel functions of the first kind, which depend on the amplitude and frequency of the AC drive, and the charges tunneling at the QPC. The tunneling current operator in the presence of the applied voltage is
\begin{equation}
I_T(t) = \sum_{\alpha} i \charge \ampl e^{-i\omega_{\alpha}(t)} \psi^{\dagger d}_{\alpha}(0,t)\psi^{u}_{\alpha} (0,t) + \text{H.c.}
    \label{eq:tunnel_current}
\end{equation}

\emph{Photo-assisted shot noise.--} Our goal is to extract unambiguous information about the fractional charge of anyons from the transport properties in the setup of Fig.~\ref{fig:Setup}. To this end, we consider the PASN which is defined as 
\begin{equation}
    \left<\bar{S} \right> = 2\int_{-\infty}^{\infty} \!\!\! d\tau \int_0^{\bar{T}} \frac{du}{\bar{T}} ~\ang{\delta I_T^2\rnd{u + \frac{\tau}{2},u-\frac{\tau}{2}}},
\end{equation}
where $\bar{T} = 2\pi/\Omega$ is the period of the AC drive, and we introduce the current-current correlations
%\begin{equation}
%   \!\! \ang{\delta I_T^2(t,t')} =\keldysh{\delta I_T(t^-) \delta I_T(t'^+)e^{-i\int_K dt H_T(t)}}.
%\end{equation}
\begin{equation}
   \ang{\delta I_T^2(t,t')} = \left \langle\delta I_T(t) \delta I_T(t') \right\rangle , 
\end{equation}
with $\delta I(t) = I(t) - \left< I(t) \right>$. The PASN is computed at lowest order in the tunneling amplitudes $\ampl$ (more details are provided in the End Matter). As a final result, one finds
%\begin{align}
%&\left<\bar{S} \right> = \!\!\!\!\!\sum_{q=\frac{e}{3},\frac{2e}{3}}\sum_{l \in \mathbb{Z}} \mu_{q}|\bar{\Gamma}|^2{J_l^2\left(\frac{q V_2}{\Omega}\right)}\frac{S^{(q)}_{DC}\left(V_1 + \frac{l\Omega}{q}\right)}{|\Gamma_{q}|^2}\label{eq:PASN_final0}\\
%&S^{(q)}_{DC}\left(V_1\right) = |\Gamma_{q}|^2 q^2\left(2\pi T\tau_0\right)^{2\scalesing-1} \left(\Gamma(2\scalesing)\right)^{-1} \nonumber \\
%& \quad  \quad \times 4\text{cosh}\rnd{\frac{q V_1}{2T}}\left|\Gamma\left( \scalesing + i\frac{q V_1}{2\pi T}\right) \right|^2 ,\label{eq:SDC}
%\end{align}
\begin{align}
\left<\bar{S} \right> &= \sum_{q=\frac{e}{3},\frac{2e}{3}}\sum_{l \in \mathbb{Z}} \mu_{q}|\bar{\Gamma}|^2J_l^2\left(\frac{q V_2}{\Omega}\right)\frac{S^{(q)}_{DC}\left(V_1 + \frac{l\Omega}{q}\right)}{|\Gamma_{q}|^2},
\label{eq:PASN_final0}\\
S^{(q)}_{DC}\left(V_1\right) &= 4 |\Gamma_{q}|^2 q^2\left(2\pi T\tau_0\right)^{2\scalesing-1} \left[\Gamma(2\scalesing)\right]^{-1} \nonumber \\
& \quad  \quad \times \cosh\rnd{\frac{q V_1}{2T}}\left|\Gamma\left( \scalesing + i\frac{q V_1}{2\pi T}\right) \right|^2 ,\label{eq:SDC}
\end{align}
where $T$ is the system temperature. Here, we introduced the total tunneling amplitude $\bar{\Gamma}$, given by $|\bar{\Gamma}|^2 = \sum_q |\Gamma_q|^2$, the tunneling amplitude $\Gamma_q$ of a given charge $q$, defined by $|\Gamma_{2e/3}|^2 = |\Gamma_1|^2$ and $|\Gamma_{e/3}|^2 = |\Gamma_2|^2+ |\Gamma_3|^2$, and the tunneling probability of each charge: $\mu_{q} = |\Gamma_{q}|^2/|\bar{\Gamma}|^2$. While $\left|\bar{\Gamma}\right|^2$ contains the non-universal physics of tunneling at the QPC, the parameters $\mu_{q} $ are dimensionless numbers, giving the relative tunneling proportions of the corresponding charges. They satisfy the relation $\mu_{e/3} + \mu_{2e/3} = 1$. The reader is warned that, for the sake of clarity, we have switched to sums over the anyon charges $q$ rather than sums over anyon types $\alpha$.

\begin{table}
\begin{center}
\centering
\begin{tabular}{c||c|c|c|c}
Type ($\alpha$)        & $\boldsymbol{g}$                & $q$ & $\delta$ \\ \hline 
$1$ & $(\sqrt{2/3} , 0 )$                    & $2e/3$     & $2/3$    \\
$2$ & $(\sqrt{1/6} , \sqrt{1/2} )$   & $e/3$     & $2/3$    \\
$3$ & $(\sqrt{1/6} , -\sqrt{1/2} )$  & $e/3$     & $2/3$ \\
\end{tabular}    
\end{center}
\caption{Anyon quantum numbers, charges, and scaling dimensions for the anyons in $\nu =2/3$ FQHE. The anyon operators are given by $\psi_{\boldsymbol{g_\alpha}} = e^{i\sum_j g_{\alpha j }\phi_j} 
=  e^{i {\boldsymbol{g}_\alpha}.\boldsymbol{\phi}} $}
\label{table:two-thirds_parameters}
\end{table}

Eq.~\eqref{eq:PASN_final0} shows that the PASN can be expressed as a sum of DC shot noises at voltages $V_1 + l\Omega/q$ \cite{tien63}. In this picture, an anyon with charge $q$, when emitted to the lower edge, ends up in a superposition of states with energies shifted by $l\Omega/q$ from the ground state. These shifts result from the absorption or emission of photons from the AC drive of frequency $\Omega$. Each state in the superposition is thus weighted by the corresponding probability of emission/absorption, given by the coefficients $J_l^2\left(\frac{qV_2}{\Omega}\right)$. 

To isolate the purely photo-assisted contributions to the PASN, we define an excess noise $\Delta \bar{S}$ by subtracting the $l=0$ contribution~\cite{kapfer19}
\begin{equation}
    \Delta\bar{S} = \left<\bar{S} \right> - \left<\bar{S} \right>_{l=0}.
\end{equation}
By plugging in the expression for the PASN in Eq.~\eqref{eq:PASN_final0}, one finds
\begin{align}
   \Delta \bar{S}  &=  \left|\bar{\Gamma}\right|^2\sum_{q = \frac{e}{3},\frac{2e}{3}}\sum_{l\ne 0}\mu_{q} J_l^2\left(\frac{qV_2}{\Omega}\right)\frac{S^{(q)}_{DC}\left(V_1 + \frac{l\Omega}{q}\right)}{|\Gamma_q|^2}.
 \label{eq:excessPASN_twothirds}
\end{align}

\begin{figure}[t]
    \centering
    \includegraphics[width=0.95\linewidth]{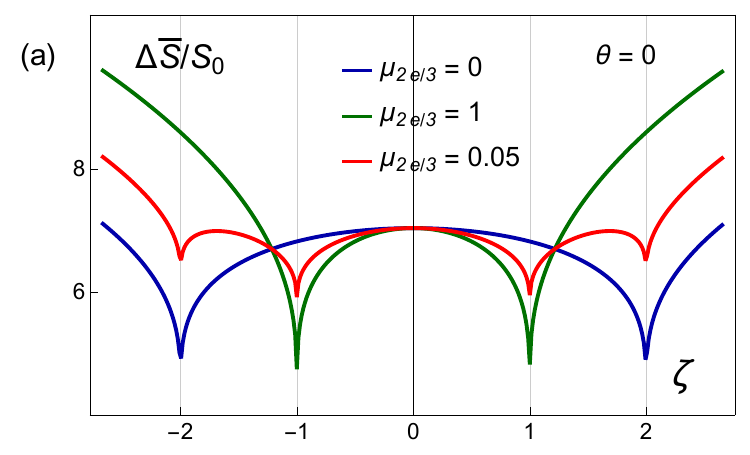}
    \includegraphics[width=0.95\linewidth]{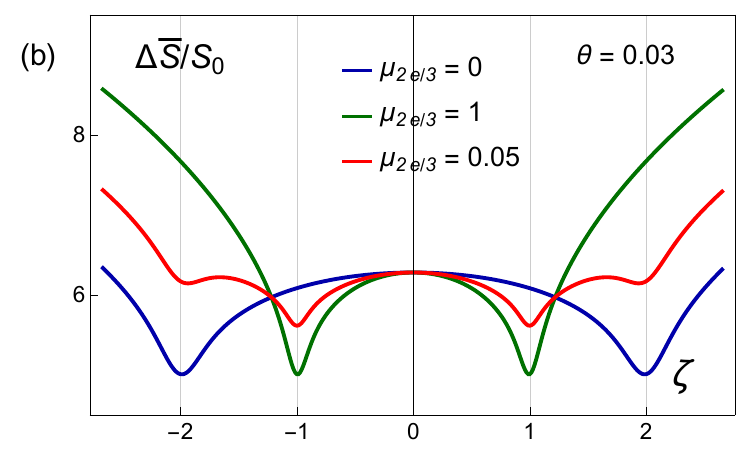}\caption{ (a-b) $\Delta \bar{S}$ as a function of $\zeta$ for different combinations of charges tunneling at the QPC. (a) $\theta = 0$, $\Delta \bar{S}$ displays dips at integer values of $\zeta$, with the position of the dips depending on the tunneling charges. When $2e/3$ charges tunnel at the QPC, $\Delta \bar{S}$ displays dips at $\zeta = \pm 1$, while when only $e/3$ charges tunnel, the dips are present at $\zeta = \pm 2$. When there is a small amount of $2e/3$ ($5\%$) tunneling along with mostly $e/3$ ($95\%$),  dips are seen at both $\zeta = \pm 1$ and $\zeta = \pm 2$.  (b) $\theta = 0.03$ constitutes an easily attainable condition in experiments. Here, the features discussed above are smoothed out by the temperature but persist. The excess PASN is normalized with respect to $S_0 = 4 {\left|\bar{\Gamma}\right|^2}\tau_0(\Omega\tau_0)^{2\delta-1}\sum_q \mu_q q^2 J_1^2(qV_2/\Omega) /\Gamma\rnd{2\delta}$. For both plots, we set $e V_2 = \Omega$.}
    \label{fig:PASN_non_trivial_scale}
\end{figure}

\emph{Unambiguous detection of multiple charges.--} Now we study excess noise as a function of the dimensionless DC voltage bias $\zeta = e\nu V_1/\Omega$, where $\nu = 2/3$ is the conductance of the edge states in the absence of the QPC.
$\zeta$ corresponds to the charge injected (in units of $e$) by the DC drive within one period of the AC drive. In Fig.~\ref{fig:PASN_non_trivial_scale}(a) we show $\Delta\bar{S}$ as a function of $\zeta$ for 
the case of zero temperature (reduced temperature $\theta = T/\Omega=0$).
Several interesting features can be observed in the excess PASN. For \(\mu_{2e/3} = 0\), indicating that only \(e/3\) charges tunnel at the QPC, \(\Delta \bar{S}\) exhibits dips at \(\zeta = \pm 2\) \cite{fqhepasn1, qhepasn1,Ferraro2014}. Conversely, in the regime where \(\mu_{e/3} = 1\), corresponding to tunneling of \(2e/3\) charges, the dips occur at \(\zeta = \pm 1\). Thus, the tunneling of different charges is distinguished by dips in the excess PASN at specific values of \(\zeta\). These values, \(\zeta = \pm 1, \pm 2\), correspond to the DC voltages at which the Josephson frequency \(\Omega^{(q)}_J = q V_1\) associated with the tunneling charge $q$ is equal to the drive frequency $\Omega$.

 Interestingly, in the intermediate regime where both charges tunnel simultaneously at the QPC, the noise dips at both $\zeta = \pm 1$ and $\zeta = \pm 2 $.
This is shown on Fig.~\ref{fig:PASN_non_trivial_scale}(a). We have chosen the value $\mu_{2 e/3}=0.05$ as most experiments seem to show that the measured tunneling charge is close to $e/3$ \cite{bid09charge, veillon24}. Even for this small proportion of  charge $2e/3$ tunneling, we observe robust dips of the excess PASN at $\zeta = \pm 1$, indicating the presence of $2e/3$ charges. This is the first main result of our Letter, showing that the excess PASN can be used as a reliable probe for the Josephson frequencies of FQHE states with multiple charges.

We now analyze the PASN for a finite temperature $\theta$. In Fig. \ref{fig:PASN_non_trivial_scale}(b) we plot the quantities discussed above for $\theta =  0.03$. This corresponds to readily accessible experimental parameters: $T \sim 30 $\emph{mK}, and $\Omega \sim 20 $\emph{GHz} \cite{kapfer19}. Here, the curves at $\mu_{2e/3} = 1$ and $\mu_{2e/3} = 0$ still display features akin to the zero temperature scenario with dips at $\zeta = \pm 1$ and $\zeta = \pm 2$ respectively, although somewhat rounded by thermal effects. At $\mu_{2e/3} = 0.05$, the dips at $\zeta = \pm 1$ and $\zeta = \pm 2$ are still clearly visible. Strikingly, the dip at $\zeta = \pm 1$ (corresponding to charges $2e/3$) is the most pronounced; this is understood by looking at Eq.~\eqref{eq:excessPASN_twothirds}, where the hyperbolic cosine and Gamma functions depending on $\zeta$ are dominant for the greater charge (when $\scalesing>1/2$).

To highlight the significance of our findings, we compare them to the information provided by the Fano factor in a similar setup in the DC regime ($V_2 = 0$). The Fano factor is defined as
\begin{equation}
 F = \sum_{q}\frac{S_{DC}^{(q)}}{2 e \, \left\langle I_T\right\rangle},  
\end{equation}
where the final expression for $S_{DC}^{(q)}$ and $\left\langle I_T\right\rangle$ are shown {respectively in Eq.~\eqref{eq:SDC} and Eq.~\eqref{eq:currfin} (see End Matter)}. {Taking} the same proportion of $e/3$ and $2e/3$ charge tunneling at the QPC as above, i.e. $\mu_{e/3} = 0.95$ and $\mu_{2e/3} = 0.05$, one has $F\simeq 0.36$ at zero temperature. The latter is {hardly} distinguishable from $F=1/3$, for purely $e/3$ charges tunneling at the QPC, within experimental errors.
{While the Fano factor shows only a small quantitative change when the tunneling charge proportion is varied, the PASN gives \textit{qualitatively} different signals depending on the charges tunneling at the QPC,
and thus provides an unambiguous detection of the different tunneling charges.}

\begin{figure}[t]
    \centering
    \includegraphics[width=0.95\linewidth]{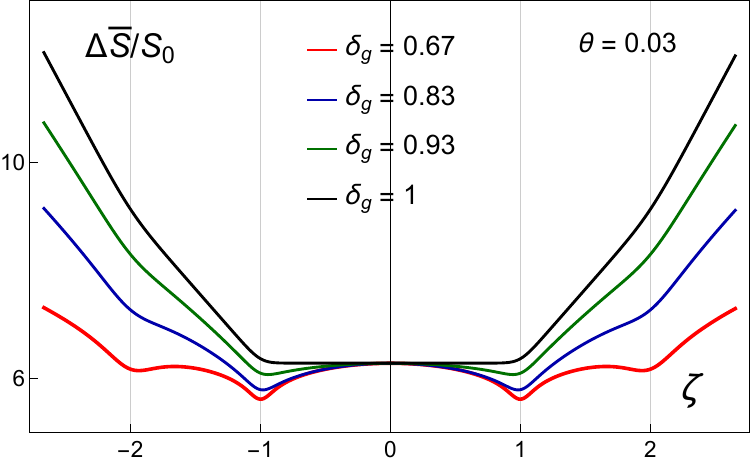}
    \caption{ Excess PASN, plotted as a function of $\zeta$ for $\mu_{2e/3} = 0.05$, and  $\theta = 0.03$ for different scaling dimensions higher than the theoretically predicted one. The PASN shows features at both $\zeta = \pm 1$ and $\zeta = \pm 2$ for all scaling dimensions. The dips at integer values of $\zeta$ for $\scalesing = 2/3$ dimension slowly morph into slope changes at the same values for $\scalesing = 1$. The features at $\zeta = \pm 2$ for $\delta = 1$ are somewhat reduced, bearing similarity to the $\mu_{2e/3} = 1$ case. The excess PASN is normalized with respect to $S_0 = 4 {\left|\bar{\Gamma}\right|^2}\tau_0(\Omega\tau_0)^{2\delta-1}\sum_q \mu_q q^2 J_1^2(qV_2/\Omega) /\Gamma\rnd{2\delta}$. We set $e V_2 = \Omega$. } 
    \label{fig:scale_PASN}
\end{figure}

It has been shown that sample and geometry-dependent details can renormalize the scaling dimension to a larger value than the theoretical predictions \cite{papa04,Braggio2012,ruelle24,schiller24}. It is hence important to test the validity of our {results} for scaling dimensions other than the theoretically predicted ones. {In Fig. \ref{fig:scale_PASN}, we plot the excess PASN for $\mu_{2e/3} = 0.05$ for different scaling dimensions between $2/3$ and $1$, for the reduced temperature $\theta = 0.03$.
%At zero temperature, shown in Fig. \ref{fig:scale_PASN}(a), the features discussed so far persist even when the scaling dimension is larger. The dips in the PASN at $\zeta = \pm 1,~\pm 2 $ when $\scalesing = 2/3$, slowly morph into slope changes at the same values when $\scalesing = 1$. Nevertheless, the PASN is sensitive to both $e/3$ and $2e/3$ charges tunneling at the QPC. As the temperature is increased, this conclusion does not change  significantly.
The features discussed so far persist even when the scaling dimension gets larger. The dips in the PASN at $\zeta = \pm 1,~\pm 2 $ when $\scalesing = 2/3$, slowly morph into slope changes at the same values when $\scalesing = 1$. Nevertheless, the PASN is sensitive to both $e/3$ and $2e/3$ charges tunneling at the QPC.}
 %the features in the PASN persist, while the minima are rounded compared to the zero-temperature case. The features at $\zeta = \pm 2$ is significantly reduced only for $\scalesing \sim 1$, but it persists for smaller values of the scaling dimension even much higher compared to the theoretical value $\scalesing = 1/3$. The PASN at higher scaling dimensions is hence largely consistent with tunneling only of $2e/3$ quasiparticles, even when $\mu_{2e/3} = 0.05$.
 This is the second main result of our Letter, {as it demonstrates that} our proposal is a reliable way to detect the topological fractional charges independently of the non-universal details of the system.  

\emph{Discussion and conclusion}.-- In this Letter, we investigated photo-assisted shot noise (PASN) as a robust probe for detecting multiple fractional charges in the fractional quantum Hall effect (FQHE) at filling factor $\nu = 2/3$. We developed a general theoretical formalism to compute PASN for Abelian quantum Hall edge states, which is presented in detail in the End Matter. Our study considered a setup with an AC sinusoidal drive combined with a DC voltage applied across a quantum point contact (QPC). While our analysis focuses on this configuration, the formalism is broadly applicable to other scenarios with any periodic, time-dependent voltage.

The PASN is analyzed as a function of the scaled DC voltage $\zeta = \nu e V_1 / \Omega$ while keeping the AC drive constant. At $\nu = 2/3$, theory predicts simultaneous tunneling of anyons with charges $e/3$ and $2e/3$. Our results demonstrate that PASN can reliably distinguish the Josephson frequencies associated with these two types of anyons. Notably, we predict distinctive features in the PASN as a function of $\zeta$, which are within reach of state-of-the-art experiments conducted at currently achievable temperatures and AC drive frequencies.

Even when the scaling dimensions of the anyonic operators deviate from theoretical predictions, our key results remain robust. This is particularly important in light of recent experimental findings, which suggest that scaling dimensions may vary substantially between samples \cite{ruelle24, veillon24, schiller24}. This robustness highlights the utility of PASN as a tool for probing fractional charges, especially in hierarchical states beyond the Laughlin sequence, where multiple charge types are expected to tunnel simultaneously at the QPC.

Future investigations could explore extensions of this work to minimal excitations \cite{levitov06,rech17} and Hong-Ou-Mandel-type collisions of anyons \cite{jonckheere23,ruelle24} in general Abelian FQHE systems. Additionally, applying this formalism to non-Abelian states presents an exciting direction~\cite{nayak08,Carrega2011}.

\acknowledgments{This work was carried out in the framework of the project “ANY-HALL” (Grant ANR No ANR-21-CE30-0064-03). It received support from the French government under the France 2030 investment plan, as part of the Initiative d’Excellence d’Aix-Marseille Université - A*MIDEX. We acknowledge support from the institutes IPhU (AMX-19- IET-008) and AMUtech (AMX-19-IET-01X).}

\bibliography{apssamp}
    \section*{End Matter}

\emph{General Abelian FQHE edge states.--}
A general Abelian FQHE edge is described by the Hamiltonian $H_0 = \sum_{j = 1}^{N} \frac{v_j}{4\pi} (\partial_x\phi_j)^2$, where $\phi_j$ denote the bosonic modes on the edge, $v_{j} > 0$ denote the propagation velocity of $j-$th bosonic mode . The bosonic fields satisfy commutation relations $   \left[\phi_j(x), \phi_k(y) \right] = i\pi\delta_{jk}\chi_j\text{sign}(x-y)
$, where $\chi_j$ denotes the chirality of the bosonic mode.  The edges host anyons of the form $   \psi_{\alpha} = e^{i {\boldsymbol{g}_\alpha}.\boldsymbol{\phi}}$ where vectors with $N$ components are denoted in bold. A general Abelian FQHE hosts multiple types of anyons on the edge, with the type denoted by subscript $\alpha$ (we refer to the SM of Ref.~\cite{iyer24} for further details).

We consider a FQHE bar with two edges on opposite sides coupled via a quantum point contact (QPC). In full generality, all the different types of quasiparticles that exist on the edge tunnel across the QPC. Thus, the tunneling of quasiparticles between the edges is given by the tunneling Hamiltonian $H_T(t) = \sum_{\alpha }  \ampl {e^{-i\omega_{\alpha}(t)}} \psi^{\dagger d}_{\alpha}(0,t)\psi^{u}_{\alpha} (0,t)+ H.c.$
 where $\ampl$ is the tunneling amplitude of type$-\alpha$ quasiparticle, {$e^{-i\omega_{\alpha}(t)}$} is the phase gathered by a type$-\alpha$ quasiparticle as it tunnels across the QPC, and the superscripts $u/d$ denote operators on the upper/lower edge. The commutator of the tunneling Hamiltonian with the charge density on the lower edge gives the current tunneling from the upper edge to the lower edge:
\begin{equation}
    I_T(t) = \sum_{\alpha} i \charge \ampl e^{-i\omega_{\alpha}(t)} \psi^{\dagger d}_{\alpha}(0,t)\psi^{u}_{\alpha} (0,t) + H.c.
    \label{eq:tunnel_current_op}
\end{equation}
where $\charge $ is the charge of the type$-\alpha$ anyon.

\emph{DC Fano factor:}
Here, we calculate the DC Fano factor in a general Abelian quantum Hall system. We start by calculating the current-current correlations at the output of the QPC. This quantity is defined within the Keldysh formalism as
\begin{equation}
 \ang{\delta I_T^2(t,t')} =\keldysh{\delta I_T(t^-) \delta I_T(t'^+)e^{-i\int_K {dt_1 H_T(t_1)}}}
\end{equation}
where $\delta I(t) = I(t) - \left< I(t) \right>$, the superscripts over time arguments indicate the position of the time arguments on the Keldysh contour, $T_K$ denotes time-ordering on the Keldysh contour, and $\int_K$ denotes an integral over the Keldysh contour. Working at lowest order in the tunneling amplitudes $\ampl$, the current-current correlations can be shown to take the form:
\begin{equation}
 \ang{\delta I_T^2(t,t')} = \keldysh{ I_T(t^-)  I_T(t'^+)}.
 \label{eq:correlations_defn}
\end{equation}
Plugging in the expression for the tunneling current operator from {Eq.~\eqref{eq:tunnel_current_op}}, the current correlation due to a constant DC voltage $V_1$ is given by
\begin{equation}
 \ang{\delta I_T^2(t,t')} = \sum_{\alpha}\charge^2|\Gamma_{\alpha}|^2 \cos{\left[\charge V_{1}(t-t') \right]} e^{2\scale\mathcal{G}(t-t')}. \\
\end{equation}
Here, we have used the relation $\omega_{\alpha}(t) = q_\alpha V_1 t$, valid for a constant DC voltage $V_1$. Moreover, we have taken the upper and lower edges to be independent, employing the correlation functions:
\begin{equation}
\begin{split}
&\left\langle 
T_K \left\{
\psi_{\alpha}^{u/d \dagger}(0, t^\eta) \psi_{\alpha}^{u/d}(0, t'^{\eta'}) 
\right\}\right\rangle  \\
= &\left\langle 
T_K \left\{ {
\psi_{\alpha}^{u/d}(0, t^\eta) \psi_{\alpha}^{u/d \dagger}(0, t'^{\eta'}) }
\right\}\right\rangle = e^{\delta_{\alpha}\mathcal{G}^{\eta \eta'}(t - t')},    
\end{split}
\end{equation}
where {$\mathcal{G}^{\eta \eta'}(t - t')$} is the Keldysh Green's function of upper/lower bosonic fields. At finite temperature $T$
\begin{equation}
\mathcal{G}^{\eta \eta'}(\tau) = 
\ln \left( 
\frac{\sinh\left(i \pi \tau_0 {T} \right)}{\sinh\left[ \pi {T} \left(i \tau_0 + \sigma^{\tau}_{\eta \eta'}\tau\right) \right]} \right),
\label{eq:B5}
\end{equation}
where \(\tau_0\) is a temporal UV cutoff of the bosonic field, and 
\begin{equation}
\sigma^{\tau}_{\eta \eta'} = 
\frac{1}{2} 
\left[ (\eta' - \eta) + \mathrm{sign}(\tau)(\eta' + \eta) \right]
\label{eq:B6}
\end{equation}
accounts for the Keldysh time ordering of the fields. The DC shot noise is then defined as
\begin{equation}
    S_{DC} = 2\int d(t-t')  \ang{\delta I_T^2(t,t')}.
\end{equation}
which, after computing the integral, gives the following expression:
\begin{align}
 S_{DC} &= \sum_\alpha 4\left|\Gamma_{\alpha}\right|^2 q_{\alpha}^2 \left(2\pi T\tau_0\right)^{2\delta_{\alpha}-1} \left[\Gamma(2\delta_{\alpha}) \right]^{-1} \nonumber \\
& \qquad \times \text{cosh}\rnd{\frac{\charge V_1}{2T}}\left|\Gamma\left( \scalesing_{\alpha} + i\frac{\charge V_1}{2\pi T}\right) \right|^2.
\end{align}
Notice that the total noise is given by the sum of noise due to tunneling of all the types of anyons existing on the FQHE edges. 

Now we proceed to calculate the average tunneling current at the QPC, defined as
\begin{equation}
   \!\! \ang{I_T(t)} =\keldysh{I_T(t^-) e^{-i\int_K dt H_T(t)}}
\end{equation}
At the lowest order in the tunneling amplitudes, it takes the form 
\begin{equation}
   \!\! \ang{I_T(t)} =-i\sum_{\eta}\eta \int dt' \keldysh{I_T(t^-) H_T(t'^{\eta})}.
\end{equation}
Plugging in the expressions of the tunneling current and tunneling Hamiltonian, following the same procedure as for the shot noise, we arrive at the following expression for the tunneling current:
\begin{align}
\langle I_T \rangle &= \sum_{\alpha} 2q_{\alpha} |\Gamma_{\alpha}|^2 \tau_0 (2\pi T \tau_0)^{2 \delta_{\alpha}-1} \left[\Gamma\left(2\delta_{\alpha}\right) \right]^{-1} \nonumber\\
& \qquad \times\sinh\left( \frac{q_{\alpha} V_1}{2T} \right)
\left|\Gamma \left(\delta_{\alpha} + i \frac{q_{\alpha} V_1}{2\pi T}\right) \right|^2, 
\label{eq:currfin}
\end{align}
where we notice again that the current is a sum over currents due to each type of anyon tunneling through the QPC.

We now define the Fano factor as the ratio of the DC shot noise and the tunneling current:
\begin{equation}
    F = \frac{S_{DC}}{2 {e} \ang{I_T}}
\end{equation}
In a general Abelian FQHE system, where different types of anyons tunnel simultaneously, this quantity is a ratio of two sums, making it difficult to extract information about the tunneling anyons. To illustrate the difficulty, we consider the large voltage limit $q_\alpha V_1 \gg T$ of the Fano factor, giving us
\begin{equation}
    F = 2\frac{\sum_\alpha |\Gamma_{\alpha}|^2 q_\alpha^{2\delta_\alpha +1 }}{\sum_\alpha |\Gamma_{\alpha}|^2 q_\alpha^{2\delta_\alpha}}.
\end{equation}
We observe that the Fano factor in this case depends on non-universal details such as the value of the scaling dimension, and the tunneling amplitudes. This is in stark contrast with the case where a unique charge $q$ tunnels across the QPC. In the latter case, the tunneling amplitude and scaling dimension drop out: $F = 2q$, {directly yielding} the charge of the tunneling anyon. 
\iffalse
To summarize, although DC Fano factor directly gives the charge of anyons when a unique anyon tunnels through the QPC (as is true in Laughlin fractions), the utility of DC Fano factor in non-Laughlin fractions is questionable. Indeed, in the latter systems, anyons with different charges can simultaneously at the QPC. In this case, the DC Fano factor depends on a non-universal average of all the tunneling charges, rendering it an inadequate probe of topologically quantized quantities.
\fi

\emph{Floquet formalism.--}We now consider driving the upper edge with a time-dependent, periodic voltage  $V(t) = V_1 + V_2(t)$ as shown in Fig.~\ref{fig:PASN_non_trivial_scale}(a), such that  $\int_{0}^{\bar{T}} dt V_2(t) = 0$, where $\bar{T}$ is the period of the AC drive. The phase gathered by the type$-\alpha$ anyons tunneling at the QPC can be shown to be $\omega_{\alpha}(t) = \charge\int_{-\infty}^{t}dt'~V(t')$. The DC voltage $V_1$ contributes a phase $e^{-i\charge V_1 t}$, and we emphasize that the voltages are fully treated by including them in the tunneling amplitude. The phase due to the periodic AC voltage is Fourier decomposed {as}
\begin{equation}
    \exp{\left(-i\charge\!\!\int_{-\infty}^t\!\!\!dt'~V_{2}(t')\right)} = \sum_l p_{\alpha}^{(l)}\!\rnd{\frac{\charge V_2}{\Omega}}e^{-il \Omega t}
    \label{eq:floquet_def}
\end{equation}
where $V_2$ denotes the amplitude of the AC drive $V_2(t)$. The Floquet components $p_{\alpha}^{(l)}(\charge V_2/\Omega)$ depend on the shape, amplitude and frequency of the periodic AC voltage applied, and the charges tunneling at the QPC. In what follows, we suppress the functional dependence of the Floquet components for brevity. The tunneling current operator can then be expressed as:
\begin{equation}
     I_T(t)  =  \sum_{\alpha, l}  i\charge \Gamma_{\alpha}p_{\alpha}^{(l)} e^{-i(\charge V_1 +  l\Omega )t} \psi_{\alpha}^{d\dagger}(0,t) \psi_{\alpha}^{u}(0,t) + {\text{H.c.}}
    \label{eq:tunnel_current_floquet}   
\end{equation}
where $\Omega = 2\pi/\bar{T}$ is the frequency of the AC drive. 

\emph{Photo-assisted shot noise.--}  We start again by calculating the current-current correlations due to the periodic drive. This amounts to plugging in the tunneling current operator of Eq.~\eqref{eq:tunnel_current_floquet} into Eq.~\eqref{eq:correlations_defn}, giving us after employing the Keldysh correlation functions:
\begin{align}
 \ang{\delta I_T^2(t,t')} &= \sum_{\alpha}\sum_{lm} \charge^2|\Gamma_{\alpha}|^2 e^{2\scale\mathcal{G}(t-t')} \nonumber \\
  & \times {\left[ p_{\alpha}^{(l)*}p_{\alpha}^{(m)} e^{i\charge V_{1}(t-t')} e^{i\Omega(lt - mt')} + 
 \text{H.c.} \right].}
\end{align}
The photo-assisted shot noise (PASN) is then defined as:
\begin{equation}
    \left<\bar{S} \right> = 2\int d\tau \int_0^{\bar{T}} \frac{du}{\bar{T}} ~\ang{\delta I_T^2\rnd{u + \frac{\tau}{2},u-\frac{\tau}{2}}}
\end{equation}
where $\tau = t-t'$ and $u = (t+t')/2$. The noise defined above is fully time-independent, as the time dependence within a period is averaged out by the second integral. Hence, the PASN takes the following form:
\begin{align}
    \left<\bar{S} \right> &=  \sum_{\alpha}\sum_{lm}  2\charge^2|\Gamma_{\alpha}|^2 p_{\alpha}^{(l)*}p_{\alpha}^{(m)} \int_0^{\bar{T}} \frac{du}{\bar{T}}e^{i(l-m)\Omega u } \nonumber \\
    &{ \qquad \times \int d\tau e^{2\scale\mathcal{G}(\tau)} \left[ e^{i \rnd{\frac{l+m}{2}\Omega+ \charge V_{1}} \tau}+ \text{H.c.} \right]}
\end{align}
where the $u$ and $\tau$ integrals decouple. The $u$ integral yields a Kronecker delta function $\delta_{lm}$, getting rid of the sum over $m$, while the $\tau$ integral is performed using known results  \cite{gradshteyn14}. We hence end up with the following final expression for the PASN for a general Abelian FQHE:
\begin{align}
\left<\bar{S} \right> &= \sum_{\alpha}\sum_{l} \left| p_l^{(\alpha)} \right|^2~S^{(\alpha)}_{DC}\left(V_1 + \frac{l\Omega}{q_{\alpha}}\right),\label{eq:PASN_final0_b}\\
S^{(\alpha)}_{DC}\left(V_1\right) &= 4\left|\Gamma_{\alpha}\right|^2 q_{\alpha}^2 \left(2\pi T\tau_0\right)^{2\delta_{\alpha}-1} \left[\Gamma(2\delta_{\alpha})\right]^{-1}\nonumber\\
& \qquad \times \cosh\rnd{\frac{\charge V_1}{2T}}\left|\Gamma\left( \scalesing + i\frac{\charge V_1}{2\pi T}\right) \right|^2.\label{eq:SDC0}
\end{align}

\end{document}